\documentclass{elsart}
\usepackage{latexsym}
\usepackage{amsmath}
\usepackage[dvips]{graphicx}
\newcommand{\beq}{\begin{equation}}
\newcommand{\eeq}{\end{equation}}
\newcommand{\beqa}{\begin{eqnarray}}
\newcommand{\eeqa}{\end{eqnarray}}
\newcommand{\bseq}{\begin{subequations}}
\newcommand{\eseq}{\end{subequations}}
\newcommand{\bold}{\boldsymbol}
\newcommand{\wtilde}{\widetilde}

\newcommand{\trm}{\textrm}
\newcommand{\lsim}{\mathrel{\mathop{\kern 0pt \rlap
  {\raise.2ex\hbox{$<$}}}
  \lower.9ex\hbox{\kern-.190em $\sim$}}}

\begin{document}

\begin{frontmatter}

\title{Meson Screening Masses in the Interacting QCD Plasma}
\author{W.M. Alberico$^1$, A. Beraudo$^2$, A. Czerska, P. Czerski$^3$}
\centerline{and}
\author{A. Molinari$^1$} 
\address{$^1$Dipartimento di Fisica Teorica dell'Universit\`a di Torino and \\ 
  Istituto Nazionale di Fisica Nucleare, Sezione di Torino, \\ 
  via P.Giuria 1, I-10125 Torino, Italy\\
  \ \\
  $^2$ECT*,\\
  strada delle Tabarelle 286, I-38050 Villazzano (Trento), Italy\\
  \ \\
  $^3$ Institute of Nuclear Physics Polish Academy of Science,
  Krak\'ow,\\
  ul. Radzikowskiego 152, Poland}


\begin{abstract}
The meson screening mass in the pseudoscalar channel
is calculated from the momentum dependent meson
spectral function, using HTL fermionic propagators.
A careful subtraction procedure is required to get an UV finite result.
It is shown that in the whole range of temperatures explored here the HTL screening
mass stays above the non-interacting result, slowly approaching the value
$m_{\trm{scr}}=2\sqrt{\pi^2T^2+m_\infty^2}$, where $m_\infty^2$ is the HTL
asymptotic thermal quark mass.
Our analysis leads to a better understanding of the excitations of QGP 
at sufficiently large temperatures and may be of relevance 
for interpreting lattice results.
\end{abstract}
\end{frontmatter}

\section{Introduction}

This work is devoted to the study of mesonic screening masses, which can be defined as
the inverse screening length characterizing the exponential falloff of the mesonic 
spatial correlator. The importance of their evaluation  in order 
to identify the relevant degrees of freedom in hot and dense QCD was first 
pointed out in~\cite{DeTar1,DeTar2,got}, where the screening masses of mesons 
and nucleons were calculated on the lattice. More recent  results 
can be found in~\cite{mtc,taro1,taro2,wis,gavai1,gavai2}.

Though obtained in different discretization schemes, 
some common features characterize the results of all the above studies.
At high temperature chiral simmetry appears to be restored: 
$\pi$ and $\sigma$, $\rho$ and $a_1$, $N_+$ (even-parity nucleon) 
and $N_-$ (odd-parity nucleon) turn out to be degenerate.
As the temperature increases meson and baryon screening masses approach
their ideal gas value of $2\pi T$ and $3\pi T$, respectively.
Finally, while the screening mass of the vector mesons approaches the 
non-interacting result immediately above the phase transition 
(hence, the presence of resonant states in this channel seems to be ruled out),
 the result for the $\pi$ and $\sigma$ mesons stays  below 
the ideal gas value up to { higher} temperatures.\footnote
{Indeed in~\cite{gavai2} it was pointed out that, in order to get reliable
results for pion properties on the lattice, one has to employ a fermionic action
which preserves chiral simmetry.}\\ 
Thus, $\pi$ and $\sigma$ states could possibly survive in the deconfined phase
 as collective excitations and be identified with the 
\emph{soft modes} of the chiral phase transition, 
as first proposed in~\cite{hat}.
In this connection, the presence of meson-like excitations 
above $T_c$ has been also
investigated within an effective PNJL model~\cite{hub06}.

Which are then the active degrees of freedom in the QGP phase, leading to
the correct interpretation of the data obtained both in the heavy-ion 
experiments and in lattice simulations? \\
For what concerns energy densities  not too far from the deconfinement transition,
the ones presently achievable in the heavy-ion collision experiments, the possible 
presence of a huge set of neutral and colored bound states surviving up to temperatures 
of order $\sim2T_c$, has been proposed ~\cite{hatsu1,hatsu2,shury03,shury}
 as being able to give a unified interpretation of
different experimental evidences: large cross sections, small mean free paths,
perfect fluid behavior, fast thermalization, elliptic flow, etc.
Actually neither the theoretical description of the QGP thermodynamics in terms of 
bound states~\cite{koch05,ratti}, nor the interpretation of the 
experimental data in terms of fast thermalization and perfect fluid
behavior~\cite{olli05} is universally accepted.

While the features of the system just above $T_c$ are still under debate, 
at higher temperatures, starting from  $T\sim 3T_c$, 
the relevant degrees of freedom should be weakly interacting quarks and gluons.
In this regime, resummation schemes based on the Hard Thermal Loop (HTL) approximation 
were developed. Different thermodynamical observables turned out to be 
well described~\cite{bla1,bla2,bla3} (in particular the slow approach 
of the entropy, { pressure and energy} density to {their} ideal gas value 
is well reproduced) in terms of resummed quark and gluon propagators. The latter imply 
a rich structure of many-body phenomena: thermal masses, collective 
excitations (plasminos and longitudinal gluons) and Landau damping.\\
It is then of interest to investigate what these dressed degrees of freedom, 
which, for large enough temperatures, nicely reproduce the QGP thermodynamics,
imply for the correlation of hadronic excitations.
    
In this paper, extending the work done in~\cite{berry,pc} where the HTL 
meson spectral functions (MSFs) at zero and finite momentum were evaluated,
we address the calculation of the $z-$axis correlator of mesonic currents.
As we will show, from the large distance behavior of the correlator one can extract 
the screening mass of the meson. We limit ourselves to the pseudoscalar channel where, as 
in~\cite{berry,pc}, the computation ``simply'' requires the convolution 
of two HTL resummed quark propagators. \\    
Though the evaluation of hadronic screening masses is a major achievement of  
lattice studies of the degrees of freedom characterizing the hot QCD 
(the large number of lattice sites available along the spatial directions allows 
one to study the large distance behavior of the correlators), analytical approaches are 
not so common in the literature.

The large distance behavior of hadronic correlators in the high temperature limit  was
first discussed in~\cite{el,shuryscr}. 
The full analytical study of the spatial mesonic correlations was first addressed 
(to our knowledge) in the non interacting case in Ref.~\cite{flor}.
Effects of the interaction on the meson screening masses were 
analyzed, for example, in Refs.~\cite{koch1,koch2,za,lai,vep} 
within a dimensional reduction framework. A positive correction of order
$g^2T$ to the non-interacting result $m^0_{\trm{scr}}=2\pi T$ was found within
this kind of approach. We will compare our numerical results to the above mentioned ones.\\
A study of the screening masses in the NJL model was presented in~\cite{flor2}, whose main 
result (beyond stressing the huge numerical difficulty of the calculation) is to show how 
in the scalar and pseudoscalar channels the value of the screening mass in the deconfined 
phase stays below the non-interacting result, thus reflecting the presence of non-trivial
correlations. This seems to agree with the lattice results
and to support the presence of pion and sigma excitations for 
temperatures slightly larger than $T_c$, representing the \emph{soft modes}
of the chiral phase transition.\\

Our paper is organized as follows.
We shortly review in Sec.~\ref{sec:mesons} the basic definitions of 
the mesonic operators, correlators and spectral functions addressed 
in this work. In Sec.~\ref{sec:HTL} we discuss how to get an UV finite
result for the $z$-axis correlator in the non-interacting case and then we present
the results obtained from the finite momentum HTL meson spectral function. 
Finally, in Sec.~\ref{sec:concl}, we summarize our results and discuss how
they compare to the ones obtained in independent approaches, both
in the continuum and on the lattice.
\section{Mesonic spatial correlation function}\label{sec:mesons}
An interesting quantity to extract informations on the properties of hot QCD
is the correlator of currents carrying the proper quantum numbers to
create/destroy mesons. In practice, in the imaginary time formalism
(which allows to study equilibrium properties in thermal field theory),
one has to compute the following thermal expectation value
\beq
{\rm G}_M(-i\tau,\bold{x})\equiv\langle\wtilde{J}_M(-i\tau,\bold{x})
\wtilde{J}_M^{\dagger}(0,\bold{0})\rangle\,,\label{eq:def}
\eeq
which provides information on how fluctuations of mesonic currents are correlated.
In the above $\tau\in[0,\beta=1/T]$, while
\beq
\wtilde{J}_M(-i\tau,\bold{x})\equiv
J_M(-i\tau,\bold{x})-\langle J_M(-i\tau,\bold{x})\rangle\;,
\eeq
denotes the fluctuation of the current operator
\beq
J_M(-i\tau,\bold{x})=\bar{q}(-i\tau,\bold{x})\Gamma_M q(-i\tau,\bold{x})\;,
\eeq
being $\Gamma_M=1,\gamma^5,\gamma^{\mu},\gamma^{\mu}\gamma^5$ 
for the scalar, pseudoscalar, vector and pseudovector channels, respectively. 

The correlator in Eq. (\ref{eq:def}) is conveniently expressed through its
Fourier components according to:
\beq
{\rm G}_M(-i\tau,\bold{x})= 
 \frac{1}{\beta}\sum_{n=-\infty}^{+\infty}
 \int\frac{d^3p}{(2\pi)^3}e^{-i\omega_n\tau}
e^{i\bold{p}\cdot\bold{x}}G_M(i\omega_n,\bold{p})\;,\label{eq:mescorr}
\eeq
where 
$\omega_n=2n\pi T$ ($n=0,\pm1,\pm2\dots$) are the bosonic Matsubara frequencies.

The correlator in Fourier space can be written in term of its spectral density
through the following representation
\beq
G_M(i\omega_n,\bold{p})=-\int\limits_{-\infty}^{+\infty}d\omega
\frac{\sigma_M(\omega,\bold{p})}
{i\omega_n-\omega} \quad \Rightarrow \quad \sigma_M(\omega,\bold{p})=
\frac{1}{\pi}\trm{Im}\,G_M(\omega+i\eta,\bold{p}),\label{eq:specrit}
\eeq
showing the link between the Meson Spectral Function (MSF) $\sigma_M$ and the
retarded correlator.

On general grounds one expects the large distance correlations
to be exponentially suppressed, the coefficient governing this damping being
identified with the mass of the (mesonic) excitation, not necessarily corresponding 
to a bound state. For example the asymptotic high temperature value $2\pi T$, already 
mentioned in the Introduction for the meson screening mass, refers to the propagation 
of a non-interacting $q\bar q$ pair, each of the two particles carrying the minimal 
fermionic Matsubara frequency $\pi T$.\\
Before proceeding in the calculations we wish to stress that the introduction of a 
thermal bath establishes a privileged reference frame. Lorentz invariance is then lost 
and the result obtained from the correlation along the $\tau$-axis (usually referred to as 
\emph{dynamical mass}) will not necessarily be equal to the one arising
from the $z$-axis correlator (\emph{screening mass}).
The latter is a quantity which is most easily extracted from lattice calculations, 
through the analysis of the exponential damping of the following correlator:
\beq
\mathcal{G}_M(z)\equiv\int\limits_0^\beta d\tau \int d\bold{x_\perp}
 {\rm G}_M(-i\tau,\bold{x_\perp},z)\;,\label{eq:gzdef}
\eeq
which for large distances is expected to display a behavior like:
\beq
\mathcal{G}_M(z)
\underset{z\to +\infty}{\sim}e^{\displaystyle{-m_\trm{scr}z}}\;.
\eeq
We notice that the integrations in Eq. (\ref{eq:gzdef}) select the components 
in Fourier space corresponding to vanishing Matsubara frequency and transverse momentum. 
Hence one can write:  
\beqa 
\mathcal{G}_M(z) & = & \int\limits_{-\infty}^{+\infty}\frac{dp_z}{2\pi}
e^{\displaystyle{i p_z z}}G_M(p_0\!=\!0,\bold{p_\perp}\!\!=\!0,p_z)
\nonumber\\
{} & = & \int\limits_{-\infty}^{+\infty}\frac{dp_z}{2\pi}
e^{\displaystyle{i p_z z}}\int_{-\infty}^{+\infty}d\omega
\frac{\sigma_M(\omega,\bold{p_\perp}\!\!=\!0,p_z)}{\omega}\;,\label{eq:gzet}
\eeqa
where in the second line use has been made of the spectral representation
given in Eq. (\ref{eq:specrit}). This will be the starting point for the present
investigation of  $z$-axis correlations.

The high-energy behavior of the spectral function, 
$\sigma_M(\omega)\sim\omega^2$, for $\omega\to\infty$ makes the integration
over $\omega$ in Eq.~(\ref{eq:gzet}) UV divergent. This problem is already
present at the level of the non-interacting theory and has to be cured 
through a proper subtraction procedure. In the next section we show how we
have solved this problem when quark and antiquarks
propagating in the QGP are described by HTL propagators.
\footnote{The first complete calculation in which this difficulty was overcome
in the non-interacting case can be found in~\cite{flor}.}. 

\section{Mesonic screening masses: HTL results}\label{sec:HTL}
Before addressing the HTL calculation of the $z$-axis meson correlator,
we briefly recall the essential aspects of the non-interacting result presented in~\cite{flor}.
The free correlator is written as the sum of two terms:
\beq
\mathcal{G}^{(0)}(z)=
\mathcal{G}_{{\rm vac}}^{(0)}(z)+\mathcal{G}_{{\rm matt}}^{(0)}(z)\;.
\eeq
The first one, temperature independent, describes the process in which the external probe 
with the quantum numbers of a meson excites a $q\bar q$ pair from the vacuum.
The second one accounts for the presence of a medium (the thermal bath) which, on the one side, 
introduces a Pauli-blocking factor for the $q\bar q$ excitation; on the other side, it allows 
a second process (not possible in the vacuum), namely the absorption of the
external probe by a (anti-)quark from the thermal bath, which is then 
promoted to an unoccupied single-particle state.\\
The expression for the vacuum piece (employing Pauli-Villars regularization
in the intermediate steps) is found to be (in the case $N_f=2$):
\beq
\mathcal{G}_{{\rm vac}}^{(0)}(z)=\frac{N_c}{\pi^2z}m_0^2\,K_2(2m_0z)\;,
\eeq
where $m_0$ is the current mass of the quark and $K_2$ is the modified
Bessel function of order 2.
The matter part, in turn, reads:
\begin{multline}
\mathcal{G}_{{\rm matt}}^{(0)}(z)=\frac{N_c}{2\pi^2}
\left(\frac{1}{z^2}-\frac{1}{z}\frac{\partial}{\partial z}\right)\\
\left[-m_0\,K_1(2m_0z)+\pi T\sum_{l=-\infty}^{+\infty} 
\exp\left(-2z\sqrt{(2 l+1)^2\pi^2T^2+m_0^2}\right)\right]\;,
\end{multline}
where $K_1$ is the modified Bessel function of order 1.\\
By summing the two contributions an exact cancellation of the vacuum part with
the first term of the matter one occurs: notably, being $m_0\ll T$, this term
represents the major contribution to $\mathcal{G}_{\rm matt}$ at large distances.
One then finds:
\beq
\mathcal{G}^{(0)}(z)=\frac{N_cT}{2\pi}
\left(\frac{1}{z^2}-\frac{1}{z}\frac{\partial}{\partial z}\right)
\sum_{l=-\infty}^{+\infty} 
\exp\left(-2z\sqrt{(2 l+1)^2\pi^2T^2+m_0^2}\right)\,,\label{eq:freefull}
\eeq 
which for large distances displays the following asymptotic behavior:
\beq
\mathcal{G}^{(0)}(z)\underset{z\to\infty}{\sim}
\frac{2 N_c T}{\pi z} \sqrt{\pi^2T^2+m_0^2}\,
e^{\displaystyle{-2 z}\sqrt{\pi^2T^2+m_0^2}}.\label{eq:freeinf}
\eeq
This leads to the non-interacting result for the meson screening mass, 
which reads:
\beq
m_{\rm scr}^{(0)}=2\sqrt{\pi^2T^2+m_0^2}
\eeq
and coincides with $2\pi T$ for vanishing current quark masses.
This result arises from a dramatic cancellation
of two terms, which otherwise would completely dominate the correlator
at large distances. The possibility of analytically performing all 
the calculations in the free case allows to extract in a clean
way the exact large distance behavior of the correlator.

In the interacting case the problem is much more involved, since the meson spectral
function is obtained only numerically. 
>From Eq.~(\ref{eq:gzet}) it appears that the $z$-axis correlator can be
expressed in terms of the finite momentum MSF. In~\cite{pc} we evaluated the
latter in the HTL approximation in the pseudoscalar channel, which amounts to
a convolution of two HTL resummed quark propagators. 
In the following we will limit ourselves to this case.
Several non-trivial many-body processes turn out to contribute to the MSF and we refer 
the reader to ref.~\cite{pc} for a detailed discussion of the various terms.\\
We then write the interacting spectral function as follows:
\beqa
\sigma_M^{\rm HTL}(\omega,p_z)&=&\sigma_M^{\rm aux}(\omega,p_z)
-\left(\sigma_M^{\rm aux}(\omega,p_z)-\sigma_M^{\rm HTL}(\omega,p_z)\right)
\nonumber\\
{}&\equiv&\sigma_M^{\rm aux}(\omega,p_z)-\sigma_M^{\rm diff}(\omega,p_z)\,,
\label{eq:diff}
\eeqa
where we introduced the auxiliary MSF $\sigma_M^{\rm aux}$.
For the latter we made the following choice. We employ the expression one gets in
the free case from the convolution of two massive quark propagators,
leading for the $z$-axis correlator to the result given in 
Eq.~(\ref{eq:freefull}). For the mass of the quarks we replace the free value $m_0$ 
by the HTL asymptotic thermal quark mass $m_\infty=\sqrt{2}m_q=g(T)T/\sqrt{3}$, 
$m_q$ being the thermal gap mass appearing in the fermionic spectrum. 
Indeed, in the HTL approximation, the asymptotic thermal mass 
governs the large-momentum regime of the quark dispersion relation.
Our choice guarantees a convergent high-energy behavior for the difference
$\sigma_M^{\rm diff}(\omega,p_z)$, which makes the integration of the latter 
in Eq.~(\ref{eq:gzet}) well defined. 

\begin{figure}[!htp]
\begin{center}
\includegraphics[clip,width=\textwidth]{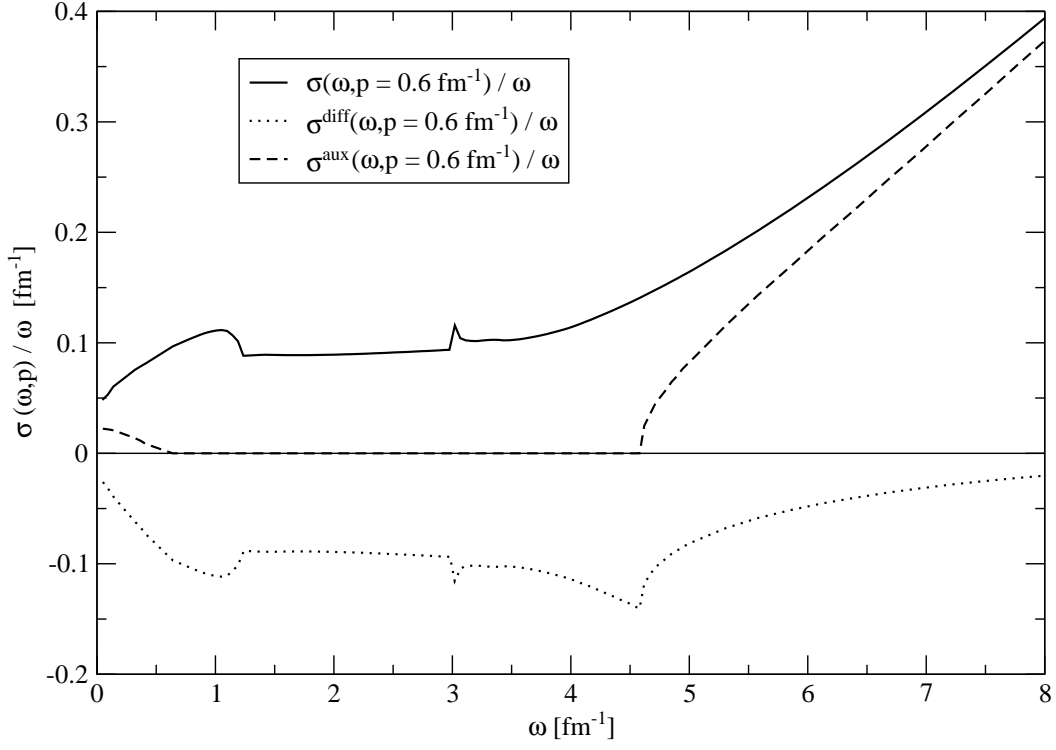}
\caption{The finite momentum pseudoscalar MSF divided by
$\omega$ together with the auxiliary one and their difference, 
for $p_z=0.6\;{\rm fm}^{-1}$ and $T=2 T_c$.}\label{p06} 
\end{center}
\end{figure}

\begin{figure}[!htp]
\begin{center}
\includegraphics[clip,width=\textwidth]{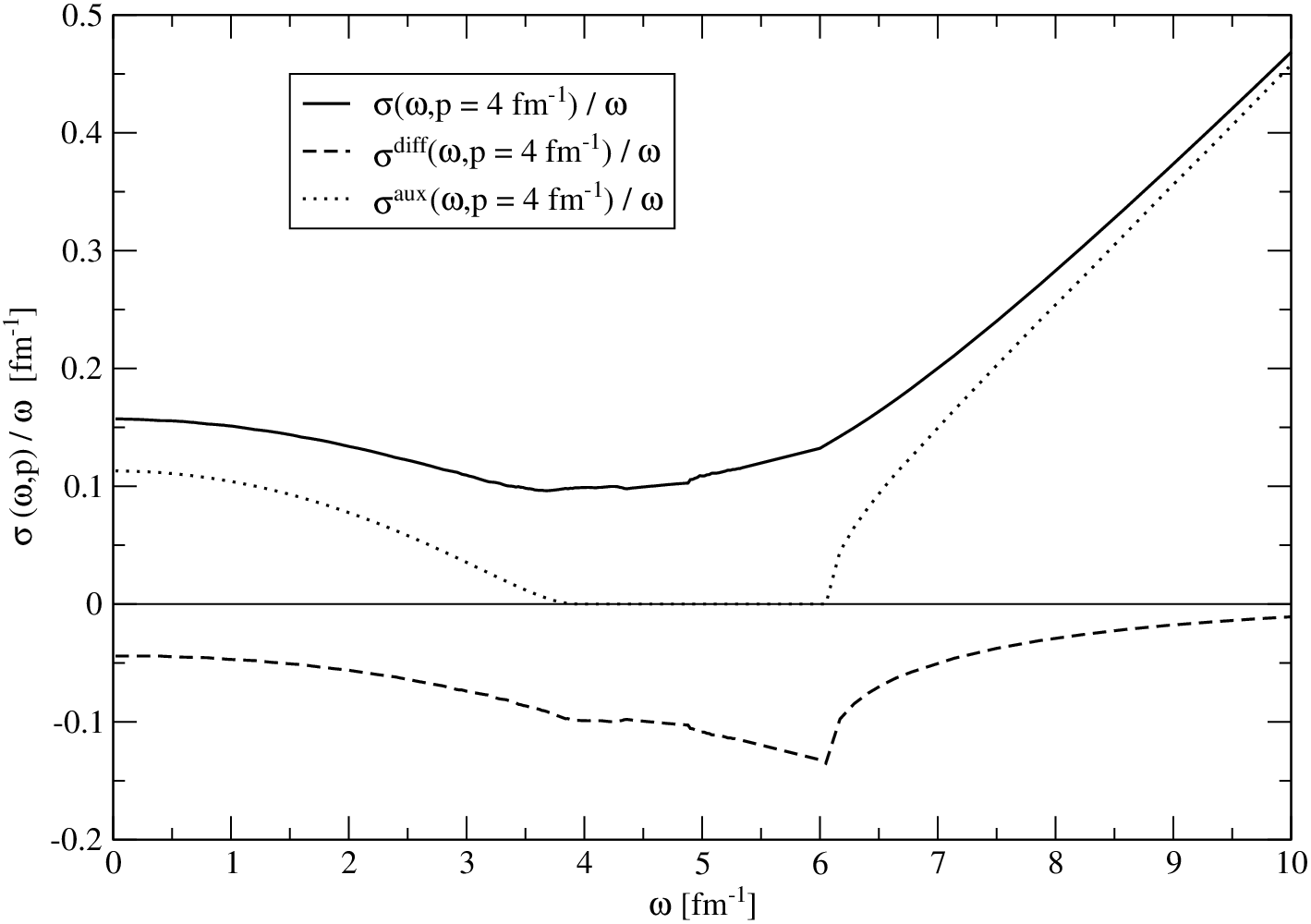}
\caption{The finite momentum pseudoscalar MSF divided by
$\omega$ together with the auxiliary one and their difference,
for $p_z=4.0\;{\rm fm}^{-1}$ and $T=2 T_c$.}\label{p4} 
\end{center}
\end{figure}

\begin{figure}[!htp]
\begin{center}
\includegraphics[clip,width=\textwidth]{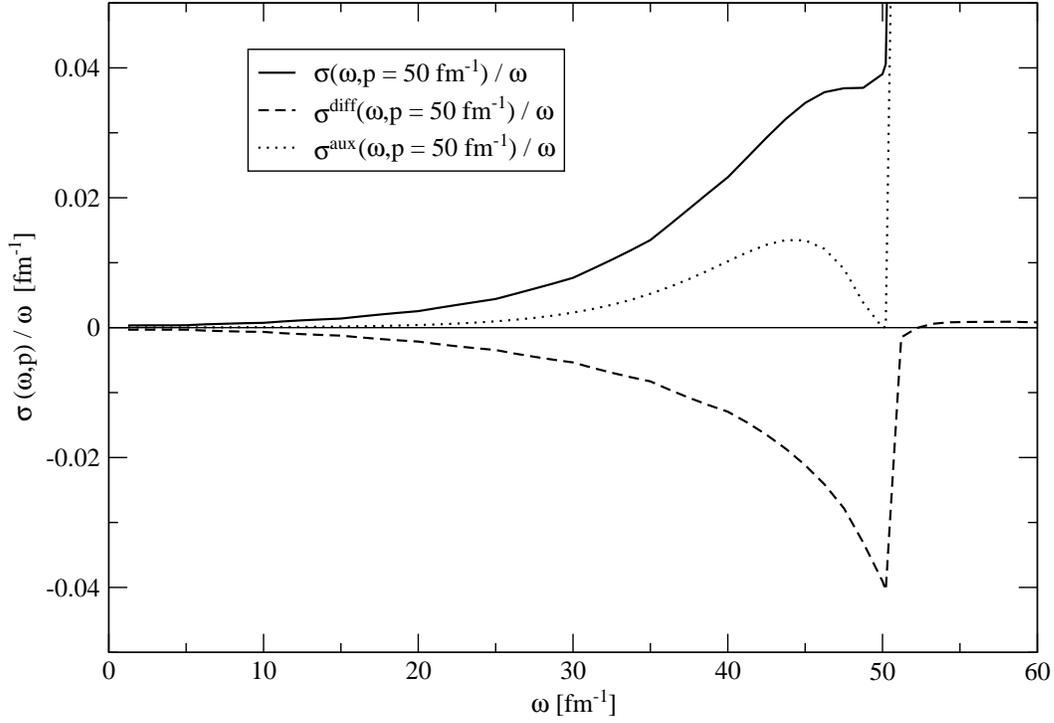}
\caption{The finite momentum pseudoscalar MSF divided by
$\omega$ together with the auxiliary one and their difference,
for $p_z=50.0\;{\rm fm}^{-1}$ and $T=2 T_c$.}\label{p50} 
\end{center}
\end{figure}

\begin{figure}[!htp]
\begin{center}
\includegraphics[clip,width=\textwidth]{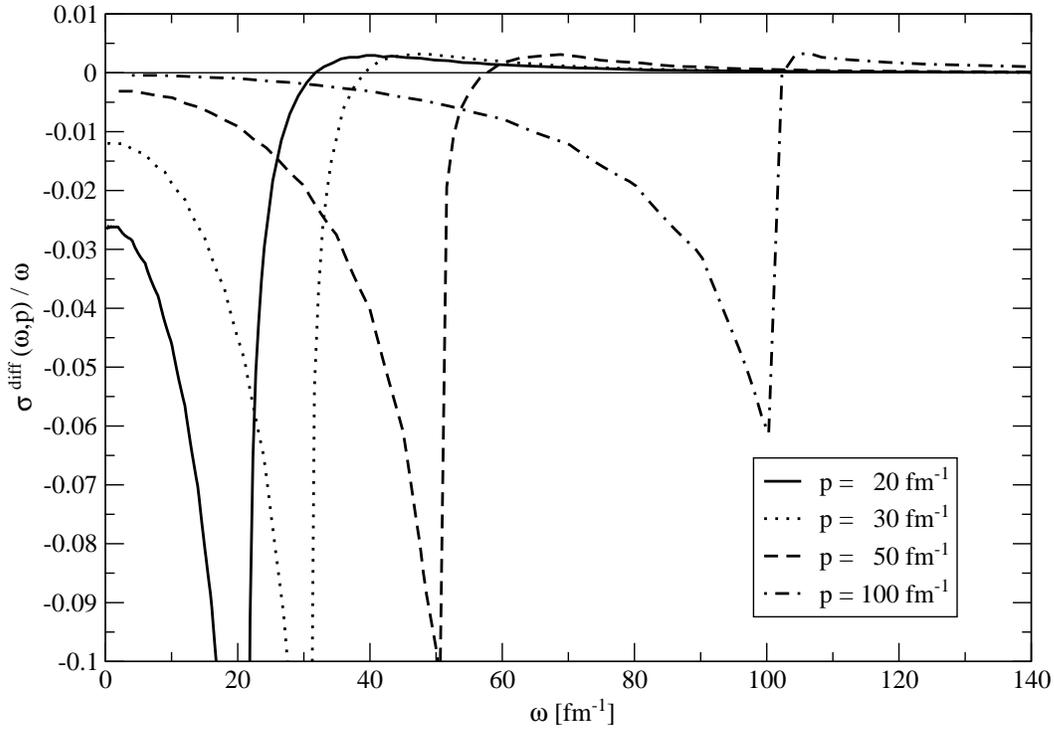}
\caption{The difference between the auxiliary and the HTL pseudoscalar MSF
 divided by
 $\omega$ for momenta $p_z=20-100
\;{\rm fm}^{-1}$ and $T=4 T_c$.}\label{diff20} 
\end{center}
\end{figure}

\begin{figure}[!htp]
\begin{center}
\includegraphics[clip,width=\textwidth]{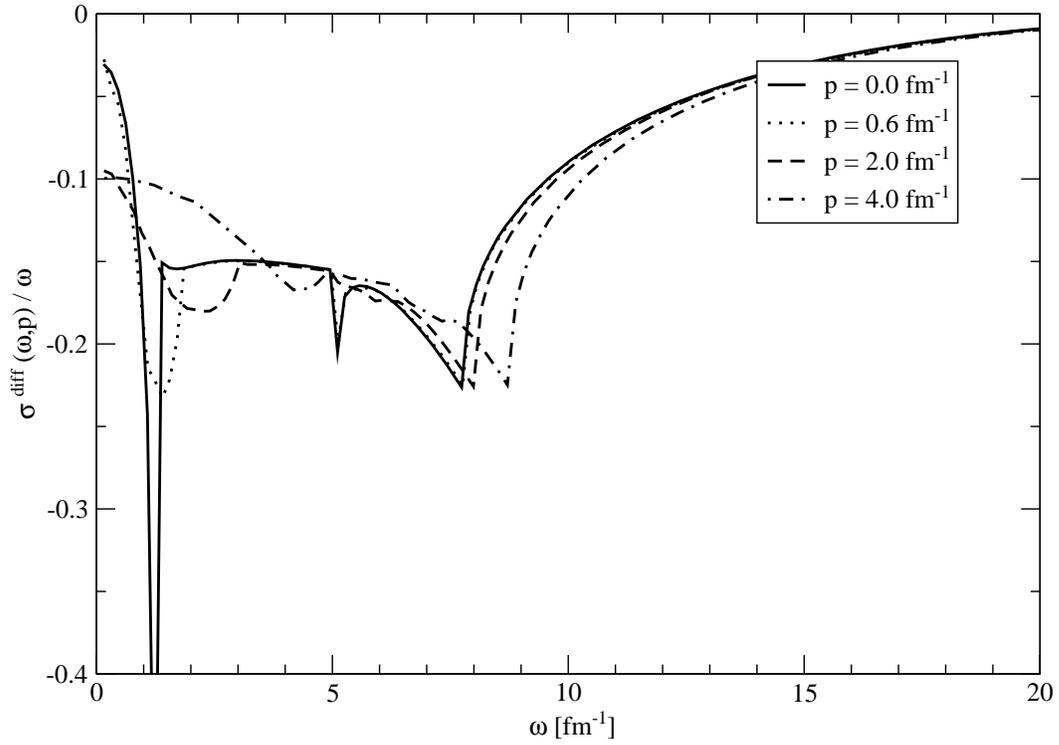}
\caption{The same as in Fig. \ref{diff20} but for momenta $p_z=0-4
\;{\rm fm}^{-1}$ and for small $\omega$.}
\label{diff4s} 
\end{center}
\end{figure}

\begin{figure}[!htp]
\begin{center}
\includegraphics[clip,width=\textwidth]{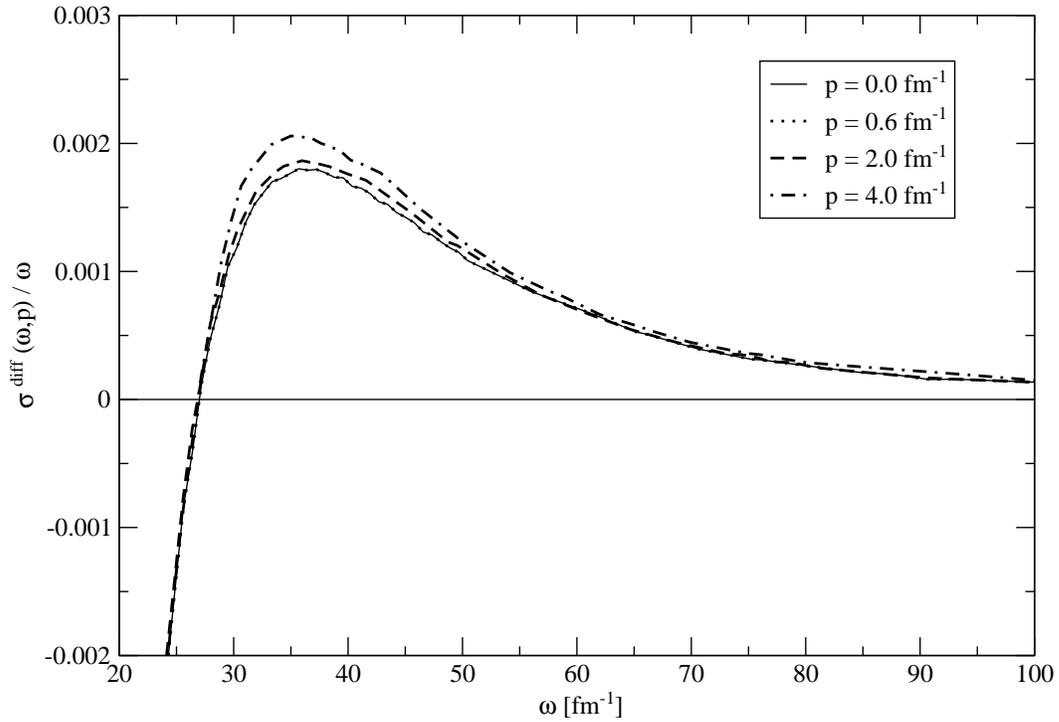}
\caption{The same as in Fig \ref{diff4s} but for large $\omega$.}
\label{diff4l} 
\end{center}
\end{figure}

In Figs.~\ref{p06}-\ref{p50},  we show the $\omega$ dependence of all the 
contributions to Eq.~\ref{eq:diff} for a few momenta $p_z$, for the case $T=2T_c$.
The numerical calculations were performed up to $\omega=2500\,{\rm fm}^{-1}$
in order to check the correct high-energy limit of the difference between
the HTL and the auxiliary spectral functions. Such a difference is well behaved and, 
as $\omega$ grows, goes smoothly to zero, as we can see in Figs.~\ref{diff20}-\ref{diff4l}.

The $z$-axis correlator accordingly reads:
\beq
\mathcal{G}^{\rm HTL}(z)=\mathcal{G}^{\rm aux}(z)-
\mathcal{G}^{\rm diff}(z)\,,\label{eq:gztot}
\eeq
where $\mathcal{G}^{\rm aux}(z)$ is given by Eq. (\ref{eq:freefull}) after
making the replacement $m_0\to m_\infty$, while $\mathcal{G}^{\rm diff}(z)$
is obtained by numerically  evaluating
\beq
\mathcal{G}^{\rm diff}(z)=\int\limits_{-\infty}^{+\infty}\frac{dp_z}{2\pi}\;
e^{\displaystyle{i p_z z}}\;G^{\rm diff}(p_z)
=\int\limits_{-\infty}^{+\infty}\frac{dp_z}{2\pi}\;
e^{\displaystyle{i p_z z}}\int_{-\infty}^{+\infty} \frac{d\omega}{\omega}\;
\sigma^{\rm diff}_{PS}(\omega,p_z)\;,\label{eq:gdiffz0}
\eeq
the integration over $\omega$ being now well defined.\\
>From Eq.~(\ref{eq:gdiffz0}) it clearly appears that the full determination
of $\mathcal{G}^{\rm diff}(z)$ would in principle require performing the
Fourier Transform (FT) of a function that we know only numerically,
after summing a huge set of different contribution which are listed 
in Ref.~\cite{pc}. In the present work, however, we are only interested in the determination of 
the coefficient which governs the exponential damping of $\mathcal{G}^{\rm diff}(z)$.
Hence we will limit ourselves to evaluate with high precision the large $z$ behaviour 
of the latter.\\
For this purpose it appears then more convenient to employ the following procedure:
we perform a fit of $G^{\rm diff}(p_z)$ with a function whose FT  in coordinate space 
is known and displays an exponential decay at large distances. For the sake of simplicity, 
for the fitting function we choose
\beq 
G^{\rm diff}(p_z) = - \sum_{i=1}^{2} \frac{2 m_i c_i}{m^2+p_z^2}
\;,\label{eq:fit}
\eeq
which leads to a $z$-axis correlator with the following functional form:
\beq 
\mathcal{G}^{\rm diff}(z) = - \sum_{i=1}^{2} c_i  e^{-m_i  z }
\;.\label{eq:gzdiff}
\eeq
The above, indeed, is a simple multi-exponential fit, which is generally used 
to extract the masses of physical states with given quantum numbers from
the study of lattice correlators.\\
We are aware that there is a priori no reason, for the above expression, to be the 
correct functional form of $\mathcal{G}^{\rm diff}(z)$, as it can be guessed, for example, 
by looking at the free result in Eq.~(\ref{eq:freefull}).
This assumption could obviously introduce some bias in our results.
Hence, while we are confident that at large distances the correlator is 
dominated by a single decreasing exponential, controlled by the smallest between 
the two parameters $m_i$, we do not attach a real physical meaning to the other coefficients
of the fitting function. 

In order to clarify our fitting procedure let us start by considering the large
distance behavior of the non-interacting $z$-axis correlator.
>From Eq.~(\ref{eq:freefull}), for the case $m_0=0$, it follows that the latter 
can be conveniently expressed in the following form:
\beqa
\mathcal{G}^{(0)}(z)&=&c_1(z)e^{-(2\pi T)z}+c_2(z)e^{-(6\pi T)z}
+\dots\nonumber\\
{}&=&c_1(z)e^{-(2\pi T)z}\left(1+\frac{c_2(z)}{c_1(z)}e^{-(4\pi T)z}
+\dots\right)\,,
\eeqa
where the $c_i(z)$ are functions displaying only a mild dependence on $z$
(compared to an exponential), which in principle can be obtained by
matching the above expression with  Eq.~(\ref{eq:freefull}).
Then one expects the $z$-dependence being dominated by the first exponential for 
distances such that
\beq
(4\pi T)\,z\gg 1\quad\Longleftrightarrow\quad z\gg\frac{1}{4\pi T}\,.
\label{eq:cond}
\eeq
We believe it to be a reasonable assumption that these considerations remain essentially
true also in the interacting case.
Then, from
\beq
\mathcal{G}^{\rm diff}(z)=\int\limits_{-\infty}^{+\infty}\frac{dp_z}{2\pi}\;
e^{\displaystyle{i p_z z}}\;G^{\rm diff}(p_z)
\eeq
it follows that in order to correctly reproduce the large $z$ behavior of
$\mathcal{G}(z)$ one needs a careful determination of $G(p_z)$ up to
$p_z\sim1/z$. Contributions from $p_z\gg1/z$ are in fact suppressed by 
{ the} oscillations of the integrand.
Hence, from Eq.~(\ref{eq:cond}), one can assume that the numerical evaluation
of the screening mass from a fit reconstruction of $G(p_z)$ requires 
a sufficiently large number of data points for $p_z\lsim 4\pi T$.
On the other hand one should not extend the fit to too large values of $p_z$,
otherwise the lighest mass would no longer control the $z$-behavior.\\
Of course a more precise determination of the range over which the fit has to be 
performed cannot be obtained through these {semi-quantitative} considerations.
In practice, in fitting  $G^{\rm diff}(p_z)$ we adopted the following procedure.
For each temperature we varied the maximum value of $p_z$ in a ``reasonable range''.
After checking that the value of the lightest mass ($m_1$) was sufficiently robust 
with respect to the arbitrary choice of $p_z^{\rm max}$ (i.e. displaying very modest
variations), for each temperature we took a weighted average of the 
different values. 

The results for $m_1$ are reported in 
Table~\ref{tab:m_1}, together with the estimate of the error resulting 
from the fitting procedure and the range over which we varied $p_z^{\rm max}$.
As a further check of the robustness of the values of $m_1$ thus obtained,
for each temperature, we extended our fit to larger values of $p_z$,
introducing a third decreasing exponential. In all the cases we found that
the values of $m_1$ obtained with this three-mass fit are indeed consistent
with the ones given in Table~\ref{tab:m_1}.
\begin{table}
\begin{center}
\begin{tabular}{|c|c|c|c|}
\hline
$T/T_c$ & $m_1\,({\rm fm}^{-1})$ & $\delta m_1\,({\rm fm}^{-1})$ 
& $p_{max}\,({\rm fm}^{-1})$\\
\hline
1 & 6.754 & 0.016 & 15$\div$50 \\
\hline
2 & 13.415 & 0.015 & 30$\div$100 \\
\hline
4 & 26.213 & 0.043 & 40$\div$200 \\
\hline
10 & 65.84 & 0.41 & 200$\div$300 \\
\hline
\end{tabular}
\caption{\label{tab:m_1} The coefficient $m_1$ extracted from the fit of
$G^{\rm diff}(p_z)$ for the different temperatures explored, together with
the range of momenta over which the fit has been performed.}
\end{center}  
\end{table}

\begin{figure}[!htp]
\begin{center}
\includegraphics[clip,width=\textwidth]{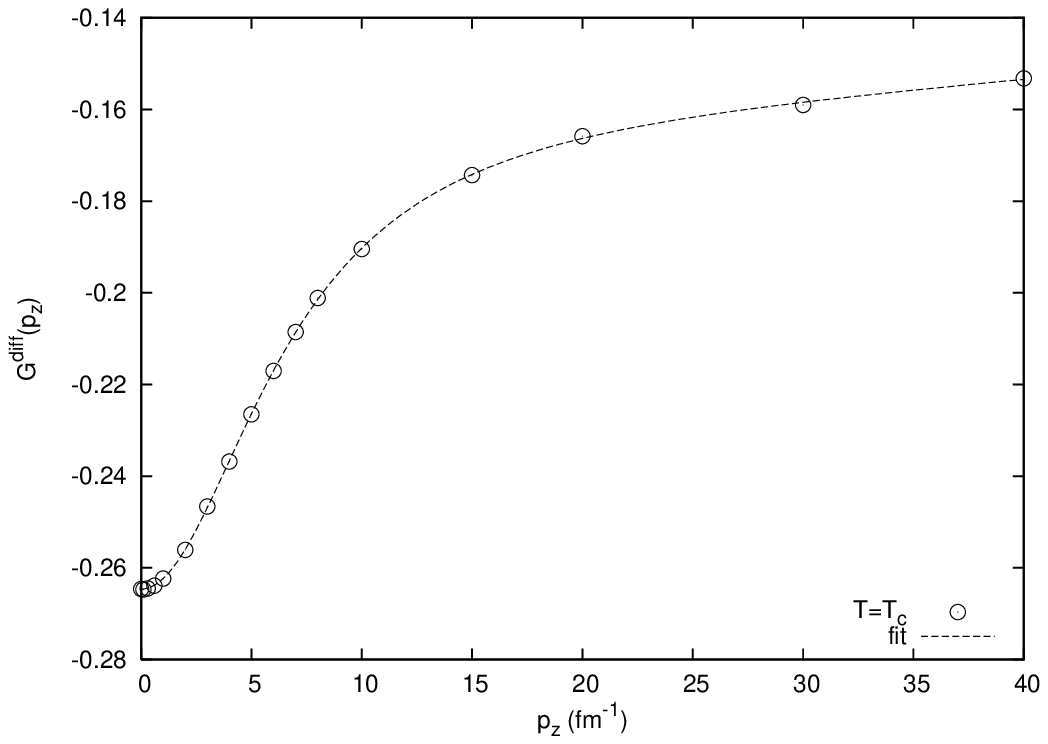}
\caption{$G^{\rm diff}(p_z)$ [in fm$^{-2}$] for the temperature $T/T_c = 1$.}
\label{gp1} 
\end{center}
\end{figure}

\begin{figure}[!htp]
\begin{center}
\includegraphics[clip,width=\textwidth]{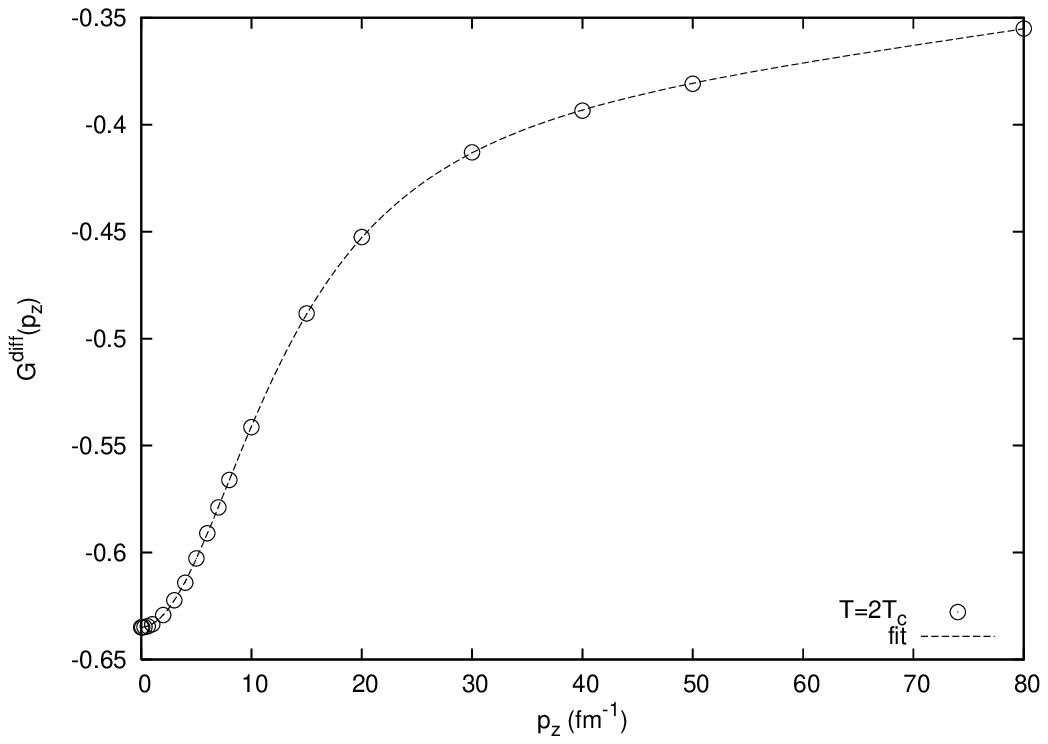}
\caption{$G^{\rm diff}(p_z)$ [in fm$^{-2}$]  for the temperature $T/T_c = 2$.}
\label{gp2} 
\end{center}
\end{figure}

\begin{figure}[!htp]
\begin{center}
\includegraphics[clip,width=\textwidth]{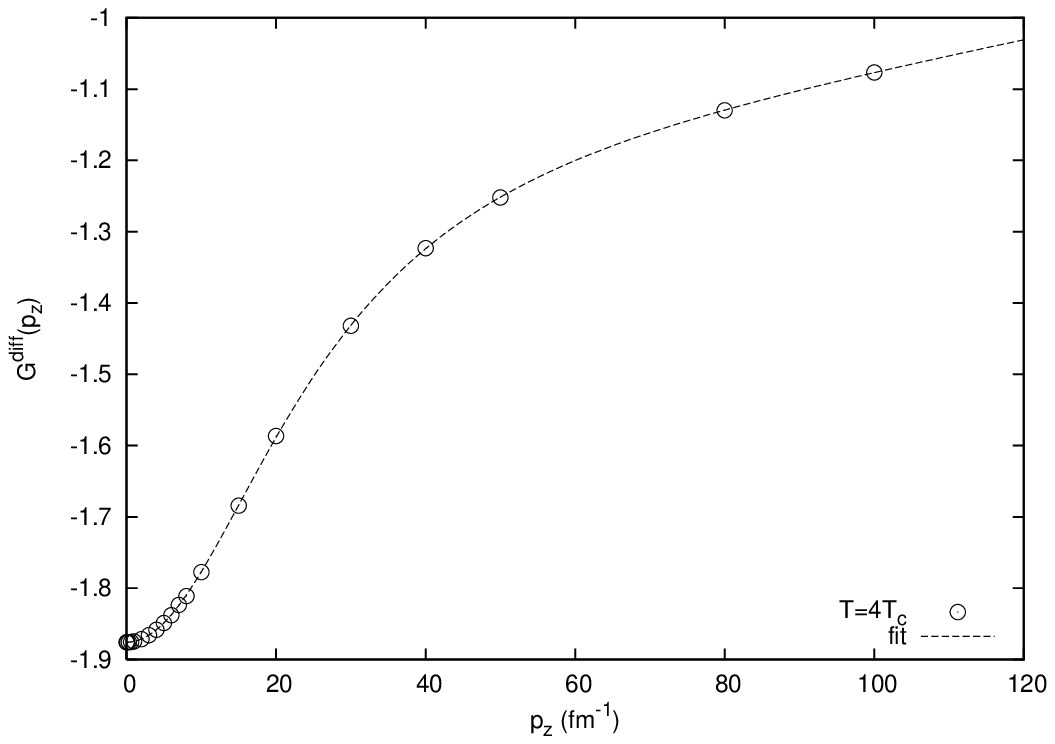}
\caption{ $G^{\rm diff}(p_z)$ [in fm$^{-2}$]  for the temperature $T/T_c = 4$.}
\label{gp4} 
\end{center}
\end{figure}

\begin{figure}[!htp]
\begin{center}
\includegraphics[clip,width=\textwidth]{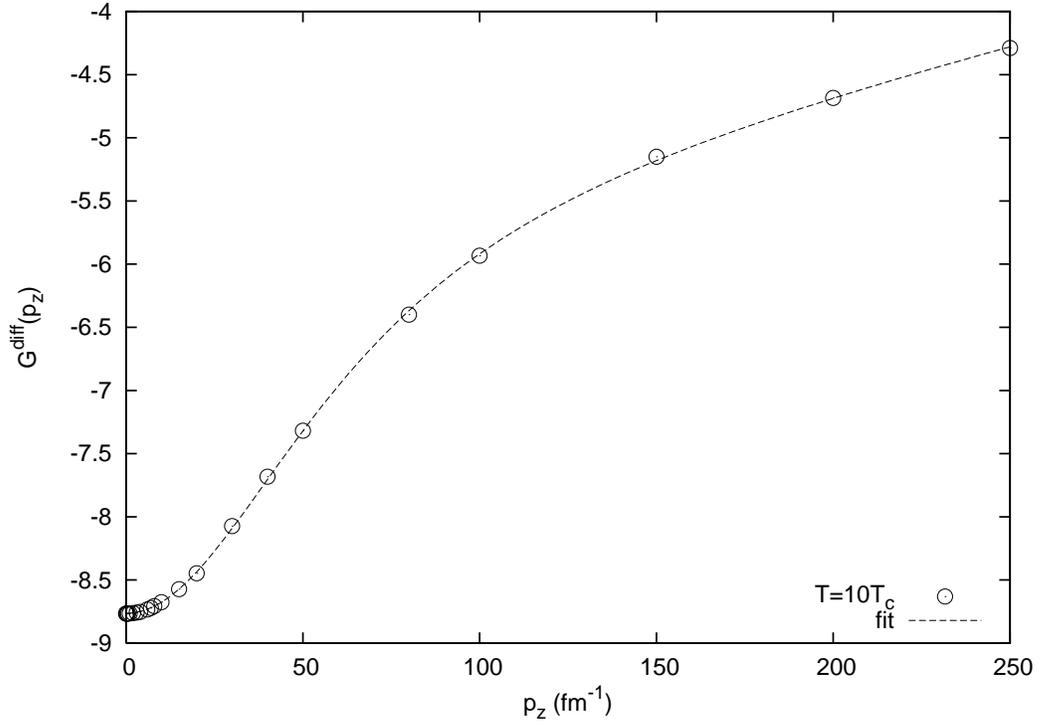}
\caption{ $G^{\rm diff}(p_z)$ [in fm$^{-2}$] for the temperature $T/T_c = 10$.}
\label{gp10} 
\end{center}
\end{figure}

The accuracy of the fit is shown in Figs.~\ref{gp1}-\ref{gp10} for different
temperatures. In all figures the numerical calculation of $G^{\rm diff}(p_z)$ 
(open circles) is compared with the smooth fitting function.

In Table~\ref{tab2} we collect our results, at different temperatures, for the 
screening mass: together with the non-interacting result ($m_{\rm free}=2\pi T$) we report 
the auxiliary HTL mass $m_{\rm{scr}}^{\rm aux}=2\sqrt{\pi^2T^2+m_\infty^2}$ 
and the lightest mass arising from our fit of ${\mathcal G}^{\rm diff}(z)$. 
Clearly the screening mass ($m_{\rm scr}^{\rm HTL}$), governing the exponential 
decay of the HTL $z$-axis correlator ${\mathcal G}^{\rm HTL}(z)$, is given by:
\beq
m_{\rm scr}^{\rm HTL}={\rm min}(m_{\rm scr}^{\rm aux},m_1)\;.
\eeq 
The Table shows that at all temperatures $m_1$ turns out to be smaller than
$m_{\rm scr}^{\rm aux}$ but larger than $m_{\rm free}$; hence it has to be assumed as the 
"true" HTL screening mass.
Notice that, independently on the result of the fit, $m_{\rm scr}^{\rm aux}$
sets an upper cutoff to the HTL screening mass. 

\begin{table}
\begin{center}
\begin{tabular}{|c|c|c|c|c|}
\hline
$T/T_c$ & $2\pi T\,({\rm fm}^{-1})$ & $m_{\rm scr}^{\rm aux}\,({\rm fm}^{-1})$ 
& $m_1\,({\rm fm}^{-1})$ & $m_{\rm scr}^{\rm HTL}\,({\rm fm}^{-1})$ \\
\hline
1 & 6.432 & 7.08 & 6.754 & 6.754\\
\hline
2 & 12.864 & 13.64 & 13.415 & 13.415\\
\hline
4 & 25.728 & 26.86 &  26.213 & 26.213\\
\hline
10 & 64.32 & 66.45 &  65.84 & 65.84\\
\hline
\end{tabular}
\caption{\label{tab2}The screening masses governing the exponential decay
of the non-interacting correlator, of the auxiliary one, of the difference
between the HTL and the auxiliary correlators, respectively. In the last column we 
show the HTL screening mass.}
\end{center}  
\end{table}

For the sake of completeness, we report in Table~\ref{tab:other} the average
values found for the other fit parameters, though we do not attach any physical 
meaning to them, in particular to the heavier mass $m_2$. Notice that, since the 
parameter $m_2$ is always very large, by limiting the range of the fit to not too large values of $p_z$ one can indeed fit equally well the data by a constant (which would correspond to a contact term in coordinate space) plus a single decreasing exponential.
We checked that in this case the value of $m_1$ results slightly larger
than the one we quote in Tables~\ref{tab:m_1} and~\ref{tab2} arising
from the two-mass fit, and approach  $m_{\rm scr}^{\rm aux}$ even faster.

\begin{table}
\begin{center}
\begin{tabular}{|c|c|c|c|}
\hline
$T/T_c$ & $c_1\pm\delta c_1\,({\rm fm}^{-3})$ &
$m_2\pm\delta m_2\,({\rm fm}^{-1})$
& $c_2\pm\delta c_2\,({\rm fm}^{-3})$\\
\hline
1 & 0.364$\pm$0.001 & 195.5$\pm$4.0 & 15.3$\pm$0.3 \\
\hline
2 & 1.754$\pm$0.003 & 291.3$\pm$1.7 & 54.4$\pm$0.3 \\
\hline
4 & 10.13$\pm$0.03 & 377.6$\pm$1.7 & 208.3$\pm$0.8 \\
\hline
10 & 126.1$\pm$1.5 & 527.4$\pm$6.6 & 1301.0$\pm$10.3 \\
\hline
\end{tabular}
\caption{\label{tab:other} The weighted average of the other fit coefficients
extracted from $G^{\rm diff}(p_z)$ for the different temperatures
explored.}
\end{center} 
\end{table} 

\begin{figure}[!htp]
\begin{center}
\includegraphics[clip,width=\textwidth]{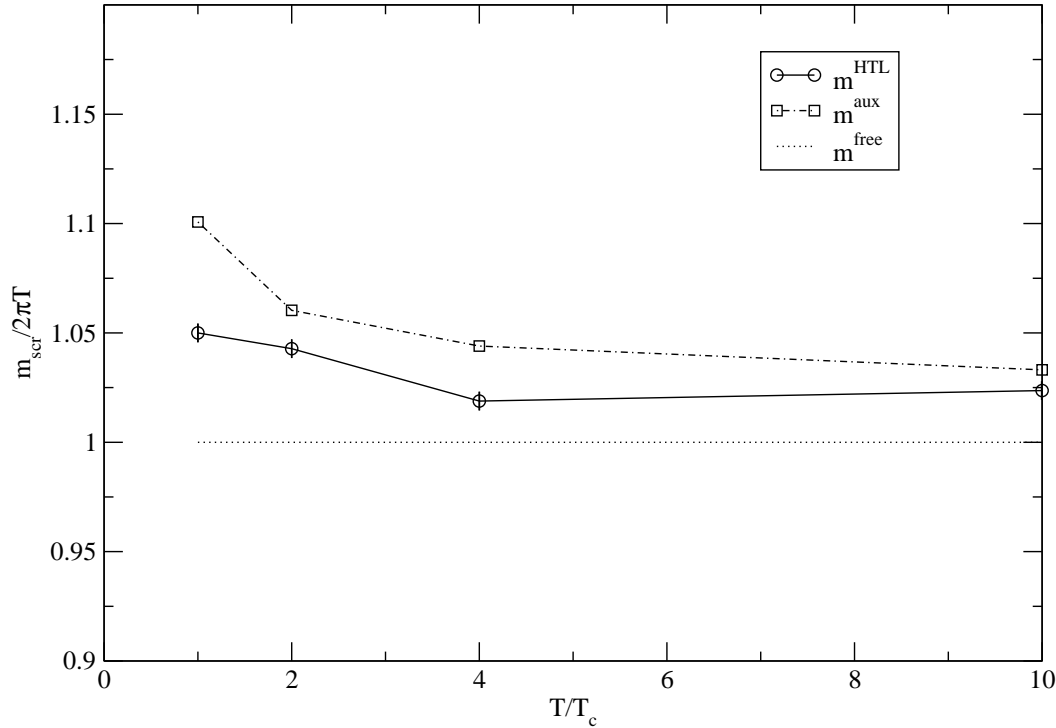}
\caption{The temperature dependence of the 
HTL result for the meson screening mass, together with
the auxiliary one, in units of the
non-interacting value $2\pi T$.}
\label{mass} 
\end{center}
\end{figure}

In Fig.~\ref{mass} we plot the HTL and auxiliary screening masses in units
of the non-interacting value $2\pi T$ as a function of the temperature.
One can see that for large temperatures, the HTL screening mass
approaches the auxiliary result
\beq
m_{\rm scr}^{\rm HTL}{\simeq}
m_{\rm scr}^{\rm aux}=2\sqrt{\pi^2T^2+m_\infty^2}\;,\label{eq:asymp}
\eeq
where $m_\infty$ is the asymptotic thermal mass of the quark. 
The high temperature behavior we have found deserves a discussion on its physical 
origin. The main effect of the interaction is to dress the quarks with
a thermal mass. Yet we remind that the HTL propagator respects chiral invariance, 
though being dominated for large momenta
by a quasiparticle pole with the dispersion relation of a massive particle.

It is indeed interesting to evaluate the lowest order correction to
the non-interacting result for the screening mass, which can be obtained by
expanding Eq. (\ref{eq:asymp}) for $m_\infty\ll\pi T$ 
(as it is indeed the case,
since $m_\infty=gT/\sqrt{3}$). One gets, for large temperatures:
\beq
m_{\rm scr}^{\rm HTL}\simeq 2\pi T+\frac{1}{3\pi}g^2T\;.
\eeq
One can compare this result with the one obtained in~\cite{lai,vep} which
can be cast in the form:
\beq
m_{\rm scr}\simeq 2\pi T+\left(\frac{1}{3\pi}+\Delta\right)g^2T\;,
\label{eq:lai}
\eeq
$\Delta$ being an additional correction term, to the same order.

Our result has been obtained through the convolution of two HTL quark
propagators, somehow assuming that the relevant degrees of freedom in the
high-temperature phase of QCD are dressed quasiparticles. This
leads (assuming the asymptotic high-temperature behavior numerically found
to be correct) to a non-perturbative result for the meson screening mass,
which implicitly contains infinite powers of $g^2$. 

In~\cite{lai,vep} the approach is completely different.
These authors do not use resummed propagators. An effective $(2+1)$ dimensional
Lagrangian is built,  in which the relevant degrees of freedom are 
the lowest {fermionic} Matsubara modes  and soft gluons. 
Within this approach the result given in Eq.~(\ref{eq:lai}) is found for the
screening mass, where the coefficient of the $g^2T$ term results from the sum
of two pieces. The first one arises from a correction to the mass of the fermionic 
modes and coincides with our result, to the extent that the HTL 
{result} is expanded 
perturbatively; the second term instead ($\Delta$) is
related to vertex and quark self-energy corrections, arising from soft gluons,
which are ignored in our approach.
{Indeed we remind the reader that the pseudoscalar vertex receives no HTL correction
and, in the evaluation of HTL self-energy diagrams, one considers the loop
integrals being dominated by hard internal momenta.}

\begin{figure}[!htp]
\begin{center}
\includegraphics[clip,width=\textwidth]{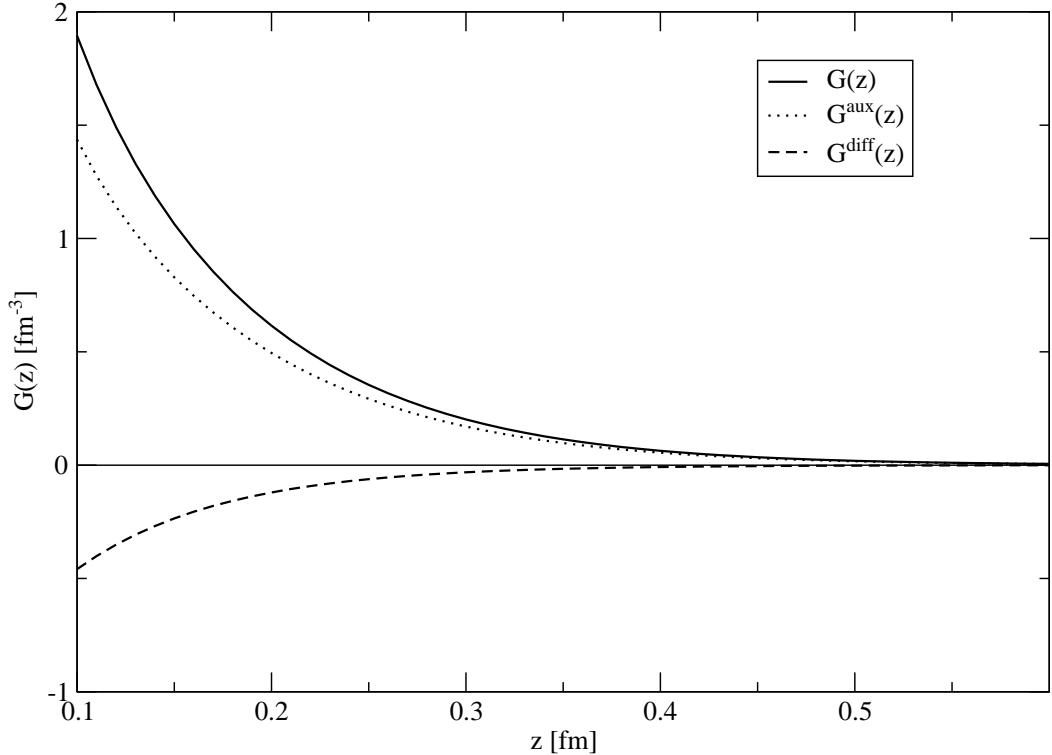}
\caption{The final result for the HTL $z$-axis correlator 
$\mathcal{G}^{\rm HTL}(z)$
(continuous line)
obtained, according to Eq. (\ref{eq:gztot}), through the difference between
the auxiliary correlator $\mathcal{G}^{\rm aux}(z)$ and 
$\mathcal{G}^{\rm diff}(z)$, which arises from our simple exponential fit.
Only the large distance behavior should be trusted. The figure refers to
the case $T=2T_c$}
\label{gz} 
\end{center}
\end{figure}
%

Both our approach and the one employed in~\cite{lai,vep}
agree in predicting a small positive correction {of order $g^2T$} to the 
non-interacting result for the meson screening mass, in contrast with the 
lattice results available so far.
The above discussed origin of the small discrepancy, between the two approximation 
schemes, which affects the exact numerical coefficient of $g^2T$, makes, in our 
opinion, this qualitative trend of the results rather robust and 
represents a challenge for lattice studies.

 Finally, in Fig.~\ref{gz} we display the behavior of the 
spatial correlator
$\mathcal{G}(z)$ for the case $T=2T_c$ obtained by combining togheter
the auxiliary correlator $\mathcal{G}^{\rm aux}(z)$ (evaluated from
Eqs.~(\ref{eq:gztot}) and (\ref{eq:freefull}) ) and the result for
$\mathcal{G}^{\rm diff}(z)$ arising from our two-mass fit.

\section{Conclusions}\label{sec:concl}
In this paper we have addressed the numerical evaluation of the screening
mass of a (pseudoscalar) meson in the QGP phase of QCD, i.e. the quantity
which governs the large-distance exponential decay of the correlations of
mesonic current operators. 

We have presented results obtained in the Hard Thermal Loop approximation.
While the pseudoscalar vertex does not receive HTL corrections, for the fermionic
lines the HTL resummed propagators have been employed.
In this scheme the spectral density of the quark correlator is 
characterized by two stable quasi-particle poles (normal quark and plasmino
mode) for time-like momenta, and by a continuum contribution in the
space-like domain related to the damping of the quark due to its interaction
with the other particles of the thermal bath (Landau damping).
In a previous publication~\cite{pc} we have shown in detail that this structure
of the fermionic propagator gives rise to an extremely rich and complex set of
many-body processes contributing to the finite-momentum spectral function
of a meson in the QCD deconfined phase.

Here we employed the previous results to derive the spatial meson correlator,
limiting the discussion to the asymptotic propagation 
along the $z$-axis and presenting numerical
results for the screening mass of a pseudoscalar meson.\\
We discussed in the text how we managed to deal with the problem of the
UV divergences (arising from the $\omega^2$ growth of the MSF at high energy)
which one encounters in the calculation: the latter are overcome through a careful
subtraction procedure which takes advantage of how this problem has been 
solved in the non-interacting case.\\
Our numerical results shows that the HTL screening mass, in the explored temperature
range, turns out to be a few percent higher than the non-interacting value
$2\pi T$, (very) slowly approaching it from above as the temperature 
increases.\\
We have also compared our results with the ones obtained in another analytical
approach based on a dimensional-reduced lagrangian~\cite{lai,vep}, which
also finds a positive correction of order $g^2T$ to the non-interacting
result. Quite interestingly, once we expand perturbatively what comes out
from our numerical findings for the asymptotic high-temperature behavior of the
screening mass, we also get a correction to the free result of order $g^2T$.
The coefficient of this correction differs from the one found in~\cite{lai,vep} 
since, in the HTL approximation, soft gluons corrections are neglected.

Concerning the lattice data available so far, to our knowledge, all
the screening masses extracted from them approach the non-interacting value 
$2\pi T$ from below. This indeed appears in conflict with the studies
performed in the continuum by us and, within a different framework, by the
authors of Ref.~\cite{lai,vep}. Whether this is related to limitations of the
present lattice results or by a breakdown of the approximations assumed
in both the continuum studies (based on a separation of the scales $T$, $gT$
and $g^2T$, which is rigorously defined only in the regime $g\ll 1$) is an open 
question.\\ 
Indeed the positive correction found
in the analytical approaches mainly arises from the thermal mass aquired by the
quark modes at finite temperature, a fact which appears sound.
On the other hand, the lattice results available so far do not extend up
to very high temperatures.
Hence it is possible that the apparent mismatch between the lattice and 
the continuum results is due to the fact that the weak coupling regime, which
makes the approximations and the separation of momentum scales justified, is
 achieved only at temperatures not yet covered by the lattice studies.  
 However a more careful investigation of the possible limitations of the
lattice results available so far might be of interest.
\section*{Acknowledgments}
One of the authors (P.C.) thanks the Department of Theoretical Physics
 of the Torino University for the warm hospitality in the 
 initial phase of this work. One of the authors (A.B.) thanks the Della Riccia
fundation for financial support and the CEA-SPhT (Saclay) for warm hospitality
during the initial part of this work. He also thanks the GGI (Arcetri) where
he spent some weeks during the final part of this work.
\appendix
\section{The running of the coupling}
Our numerical results refer to the case $N_f=2$.
The transition temperature, following \cite{kacz}, has been set to the value
$T_c=202$ MeV. The ratio $T_c/\Lambda_{\overline{MS}} = 0.7721$ for the 
  $N_f=2$ case is taken from Refs. \cite{karsch,goc,zan}.
The running of the gauge coupling is given by the two-loop perturbative 
beta-function, leading to the expression:
\beq
g^{-2}(T) = 2 b_0 \log{\frac{\mu}{\Lambda_{\overline{MS}}}} +
\frac{b_1}{b_0} \log\Big\{2 \log{\frac{\mu}{\Lambda_{\overline{MS}}}}
\Big\}\label{eq:running}
\eeq
where
$
b_0=\frac{1}{16 \pi^2} \Big( 11 - 2 \frac{N_f}{3} \Big)$, $
b_1=\frac{1}{(16 \pi^2)^2} \Big( 102 - 38 \frac{N_f}{3} \Big)
$
The renormalization scale $\mu$ is usually taken to be of the order of the
temperature, which, apart from $\Lambda_{QCD}$, represents
the only physical scale entering into the problem.
For what concerns its precise numerical value 
one should, in principle, let it vary within a reasonable range in order
to get an estimate of the theoretical uncertainty of the calculation.
Here, for the sake of simplicity, due to the huge numerical work required to
produce our results, we adopt the choice 
$\mu = 1.142 \pi T$ suggested in \cite{kacz}.

\end{document}